\documentclass[aps,prb,onecolumn]{revtex4-1}

\usepackage{hyperref}
\usepackage[utf8]{inputenc}
\usepackage{amsmath}
\usepackage{bm}
\usepackage{graphicx}
\usepackage{verbatim}
\usepackage{color}
\usepackage{subfigure}

\newcommand{\etal}{\textit{et al.}}
\newcommand{\eg}{\textit{e.g.}}
\newcommand{\ie}{\textit{i.e.}}
\newcommand{\cf}{\textit{cf.}}
\newcommand{\angstrom}{\textup{\AA}}

\begin{document}

\title{Electrochemical doping of few layer ZrNCl from first-principles:\\electronic and structural properties in field-effect configuration}
\date{\today}
\author{Thomas Brumme}
\author{Matteo Calandra}
\author{Francesco Mauri}
\affiliation{CNRS, UMR 7590, F-75005, Paris, France}
\affiliation{Sorbonne Universit\'{e}s, UPMC Univ Paris 06, IMPMC - Institut de Min\'{e}ralogie, de Physique des Mat\'{e}riaux, et de Cosmochimie, 4 place Jussieu, F-75005, Paris, France}

\begin{abstract}
We develop a first-principles theoretical approach to doping in field-effect devices.
The method allows for calculation of the electronic structure as well as
complete structural relaxation in field-effect configuration using density-functional theory.
We apply our approach to ionic-liquid-based field-effect doping of monolayer, bilayer, and trilayer
ZrNCl and analyze in detail the structural changes induced by the electric field.
We show that, contrary to what is assumed in previous experimental works,
only one ZrNCl layer is electrochemically doped and that this induces large
structural changes within the layer.
Surprisingly, despite these structural and electronic changes, the density of states at the Fermi
energy is independent of the doping.
Our findings imply a substantial revision of the phase diagram of electrochemically doped ZrNCl
and elucidate crucial differences with superconductivity in Li intercalated bulk ZrNCl.
\end{abstract}


\maketitle

\section{Introduction}
In recent years, materials with reduced dimensionality have attracted a
lot of attention because of their interesting physical properties and
proposed applications range from electronics to sensing. In addition to
graphene, which is probably the most studied two-dimensional (2D)
material, monolayers or few-layer systems (nanolayers) of layered
materials such as transition metal dichalcogenides\cite{radisavljevic2011,ye2012,yuan2013,radisavljevic2013,wu2013,das2013_1,das2013_2}
or chloronitrides\cite{ye2010,kasahara2010,schurz2011,zhangs2012,ekino2013} are gaining importance because of their intrinsic band gap.
The possibility of doping these few layer systems with field-effect transistors (FETs) is particularly appealing.
In ionic-liquid-based FETs the charging of the nanolayers is substantial and thus 
allows for electrochemical doping of few-layer materials\cite{yuan2009,ye2010,ye2012,zhang2012,taniguchi2012,yuan2013,perera2013}
with surface charges of the order of $10^{14}\:\mathrm{cm}^{-2}$.
In the field of condensed matter physics, this prospect is very appealing as it becomes
possible to dope ultrathin films of layered material and explore their phase diagram to detect
charge-density waves, spin-density waves, or superconducting
instabilities in reduced dimensionality.
Furthermore, it was recently shown that the device characteristics of FETs using an ionic-liquid
are much better than those of back-gated devices\cite{perera2013}.

An eminent example is the electrochemical doping of the semiconducting chloronitride ZrNCl\cite{ye2010}.
It has been shown by Ye \etal{} that $10-30\:\mathrm{nm}$ ZrNCl flakes
can be made first metallic and then superconducting by electrochemical doping.
A similar behavior has also been detected in MoS$_2$ flakes\cite{ye2012}.
Despite these challenging experimental perspectives, the understanding of electronic and structural
properties at high electric field in FET configuration is still limited.
For example, the effects of the large electric field (in the order of $1\:\mathrm{V/nm}$\cite{ye2010,ye2012})
on the structural properties of the nanolayer remains an open question.
Moreover, in the case of ZrNCl, the surface charge, and consequently the effective doping, changes by an order of
magnitude depending on the assumed thickness of the charge layer\cite{ye2010}.
It is then crucial to address these issues to determine the phase diagram of electrochemically doped ZrNCl.

We develop a first-principles theoretical approach to doping in field-effect devices.
The method allows for calculation of the electronic structure as well as
complete structural relaxation in field-effect configuration using density-functional theory.
We apply our approach to ionic-liquid-based field-effect doping of monolayer, bilayer, and trilayer
ZrNCl.

The paper is organized as follows: in Sec. II we give a brief introduction to field-effect doping and
a description of the model we use in order to investigate such systems from first-principles, along with
the computational details. Results for the specific example of the electrochemical doping of ZrNCl are presented in Sec. III.
First, we show the structural changes in field-effect configuration and then investigate the
distribution of the induced charge in more detail. Afterwards we use a simple model to describe the modification of the band
structure and show that, despite the structural and electronic changes, the density of states at the Fermi energy is independent
of the doping. Finally, conclusions are presented in Sec. IV.

\section{Theoretical background}
\subsection{Field-effect configuration}
\begin{figure}[b]
 \centering
 \includegraphics[width=0.35\textwidth,clip=]{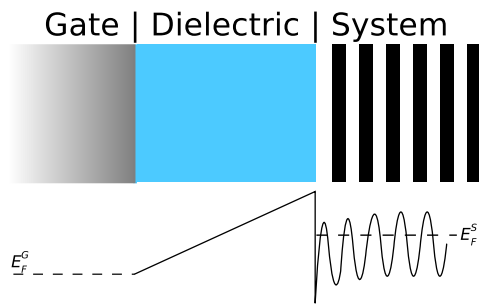}
 \caption{\label{fig:fet}Field-effect configuration in which the gate is separated from the system via a dielectric.
          The Fermi energies of the system and the gate are indicated by dashed lines and labeled $E_F^S$ and $E_F^G$, respectively.
          At the bottom the electrostatic potential of the system is sketched. The difference in the Fermi energies of
          the gate electrode and the system leads to an accumulation of charges close to the system-dielectric interface.}
\end{figure}

In Fig.~\ref{fig:fet} the typical setup for a field-effect measurement is shown. The system is separated from the
gate electrode by a dielectric such as silicon oxide.
Applying a gate voltage between the system and the gate results in a change in the Fermi level of the metal ($E_F^G$)
with respect to the one in the system ($E_F^S$). Hence, the dielectric is polarized due to the opposite charges at
the dielectric/metal and dielectric/system interfaces.

The distribution of charges in the system/dielectric/ metal-gate system generates the potential profile shown in
Fig.~\ref{fig:fet}. Inside the dielectric, a constant and finite electric field (linear potential) occurs due only
to the opposite signs of the charges on the different sides of the dielectric.
At the system/dielectric interface (on the system side), the potential varies very quickly because of the strongly
inhomogeneous charge distribution in the first few layers of the system.
Finally, inside the system and far from the dielectric, the potential felt by the electrons oscillates with a periodicity
that is related to the crystal periodicity and is determined by the ionic positions. In this region,
the spatial average of the electric field is zero. The same situation occurs inside the metal gate far from the dielectric.

In order to describe such a setup in theory, different methods have been used in the past.
In perhaps the most advanced method one takes into account the connection of the system to the gate electrode
and the back contact, effectively simulating an open-boundary system. This can be achieved by coupling the
calculation of the electronic structure of the system---often done within density-functional theory (DFT)---with the
Keldysh nonequilibrium Green's function (NEGF) formalism\cite{liu2012,gong2013}.
However, these calculations are very expensive since the electrostatic potential of the system has to be determined
self-consistently. This is achieved by solving the Poisson equation with the bulk-like Hartree potentials of the electrodes as boundary conditions
on the interfaces between the electrodes and the system. In fact, by using NEGF, one is often restricted to use the structure which
was relaxed without electric field, effectively neglecting the interplay of electric field, charging, electronic structure, and
the geometry of the system\cite{liu2012,gong2013}.
Furthermore, these simulations of the full system/dielectric/metal-gate system are not only difficult and time consuming but also
unnecessary, as we are mainly interested in what happens at the system/dielectric interface.
We thus need to simplify the modeling of the FET measurements.

The first simplification is to include the external electric field and eliminate the metallic gate.
This amounts to replacing the metallic gate with a charged plate that simulates the charge occurring
on the gated metal surface in contact with the dielectric\cite{gava2009,uchida2009}. It is important to remark that, if the
charged plate is neglected and only an external field is added, no charging can occur on the system side. 

After eliminating the gate, the next step is trying to reduce the size of the
dielectric or to eliminate it completely. Since we expect that the relevant physics occurs inside the system
in close contact with the dielectric, it is possible to replace the several nanometer thick-dielectric with a
much thinner one. This is not an approximation, but is what is found in modern nanoelectronics, where,
in order to increase the FET capacity, thinner and thinner dielectrics are used.
Finally, the ultimate approximation is the complete removal of the dielectric,
and including the charged plate close to the position of the system/dielectric interface.

The total removal of the dielectric can be achieved by following the steps mentioned above, as long as
the ions of the system are fixed. However, we also want to simulate the structural relaxation of the
system in FET configuration, which means that we need to allow the ions to move. In this case an
additional problem occurs. Having totally eliminated the dielectric, the ions can move too close to the
charged plate which simulates the gate electrode (opposite charge on the plate and the system).
This can be avoided by representing the dielectric with a constant potential barrier which
effectively mimics the physical repulsion exerted by the dielectric. This is the approximation adopted
in the present work.

Up to now, we have introduced a simpler model of the system/dielectric/gate system that works with
open boundary conditions. An additional conceptual step is needed in order to simulate the FET using
periodic boundary conditions (PBC). With PBC, the charged plate interacts with the periodic
image of the system generating a spurious electric field. This is unphysical, as in experiments
far from the dielectric and inside the system or the metal gate the average electric field must be zero. 
One way to circumvent this issue is to include a mirror image of the system that forces the electric field to
be zero between the periodic images\cite{uchida2009}. However this implies calculating a system with twice
the number of atoms and results in an higher computational load. A smarter solution comes from
electrostatics. It is indeed sufficient to add an additional dipole plate (two planes of opposite charge\cite{bengtsson1999})
generating an electric field that exactly cancels the one on the left side of the gate in Fig.~\ref{fig:fet}.
In this case no additional computational load is needed and the condition of zero electric field far from the
system/dielectric interface is enforced.

\subsection{Model}
\label{monopole}

\begin{figure}
 \centering
 \includegraphics[width=0.388\textwidth,clip=]{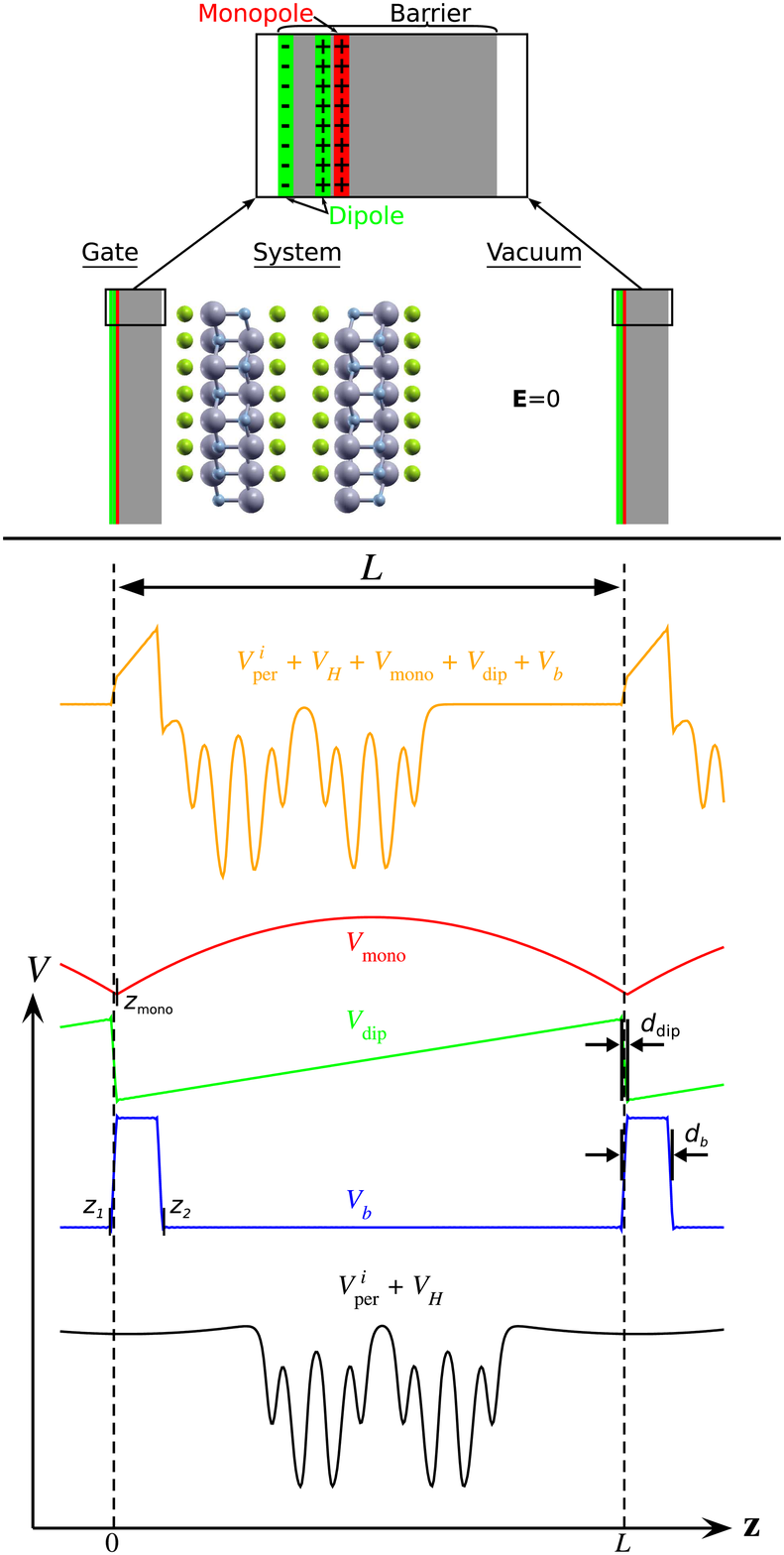}
 \caption{\label{fig:pots}(color online) Schematic picture of the planar averaged Kohn-Sham potential for periodically repeated,
          charged slabs (without the exchange-correction potential $V_\mathrm{XC}$).
          In the uppermost panel a sketch of a gated system in a periodically repeated unit cell is shown.
          Within the vacuum region between the repeated images the electric field should be zero (\ie, the total
          potential should be constant) while between the gate and the system it is different
          from zero. Due to the PBC we need to cancel the dipole of the monopole-nanolayers system so that $\mathbf{E}=0$ in
          the vacuum. This can be done by placing a dipole in the vacuum between the repeated images\cite{bengtsson1999}.
          We furthermore include a potential barrier (gray area) to mimic the dielectric.
          The different parts of the total Kohn-Sham potential are shown with different color:
          red -- monopole, $V_\mathrm{mono}$, green -- dipole, $V_\mathrm{dip}$, blue -- potential barrier, $V_b$.
          The position of the monopole is indicated by $z_\mathrm{mono}$. The length of the unit cell along $\mathbf{\hat{z}}$
          is given by $L$.
          The potential from a standard calculation using PBC is also shown (labeled $V^i_\mathrm{per}+V_H$).
          In this case, the charging leads to a finite curvature in the vacuum region due to the uniform compensating background charge.
          In principle, the vacuum region needs to be increased so that $\mathbf{E}\rightarrow0$ at the surface of the system. This in
          turn increases the depth of the potential well between the repeated images (due to the uniform compensating background charge) which can lead
          to charge spilling into the vacuum\cite{topsakal2012}.}
\end{figure}

The general setup we use to model a system in FET configuration using PBC in the three spatial directions is shown in Fig.~\ref{fig:pots}.
The nanolayers are placed in front of a charged plane\cite{gava2009,uchida2009} which models the gate as already mentioned in the
last section (henceforth referred to as ``monopole''). The layers are then charged with the same amount of opposite charge $-n_\mathrm{dop}\,A$ ($n_\mathrm{dop}$
and $A$ are the number of doped electrons per unit area and the area of the unit cell parallel to the surface respectively)
which will lead to a finite electric field in the region between the gate and the system. In the vacuum region between the
repeated images (PBC) the electric field has to be zero in order to correctly determine the changes in the electronic structure
and the geometry for such a field effect setup.
However, the dipole of the charged system and the gate leads to an artificial electric field between the different slabs of the repeated
unit cell since in PBC the electric field averaged over the unit cell is zero. We thus include an electric dipole generated by two planes of opposite
charge\cite{neugebauer1992,bengtsson1999,meyer2001} which are in the vacuum region next to the monopole (Fig.~\ref{fig:pots}).
We furthermore include a potential barrier to avoid the direct interaction between the charge density of the system and the
monopole/dipole.
The resulting total potential together with the definition of the different variables used in the following discussion is shown in Fig.~\ref{fig:pots}.

The total energy of the system in this model-FET setup is given as
\begin{align}
\label{eq:etot}
 E_\mathrm{tot} &= E^\mathrm{per}_\mathrm{tot} + E_\mathrm{mono} + E_\mathrm{dip} + E_b.
\end{align}
The first term appearing on the right-hand side of Eq.~(\ref{eq:etot}) is the total energy calculated within PBC
using the periodic, bare ion-electron potential $V^i_\mathrm{per}\left(\mathbf{r}\right)$
\begin{align}
\label{eq:eper}
 E^\mathrm{per}_\mathrm{tot} &= \sum_i\langle\psi_i\left|T\right|\psi_i\rangle+E_H+E_\mathrm{XC}\notag\\
                             &\quad+E_\mathrm{i-i}+\int_\Omega n\left(\mathbf{r}\right)V^i_\mathrm{per}\left(\mathbf{r}\right)d^3r.
\end{align}
Here the first four terms represent the kinetic energy, Hartree energy, exchange-correlation energy, and Madelung energy,
respectively, while the last term is the bare ion-electron energy with $n\left(\mathbf{r}\right)$ being the electron density.
$E_\mathrm{mono}$ is the energy associated with the monopole
while $E_\mathrm{dip}$ is the energy due to the dipole.
The interaction with the potential barrier which is used to mimic the dielectric is given by the last term in Eq.~(\ref{eq:etot}), $E_b$.

The total charge density in the cell $\Omega$ is the sum of the electronic part, the ionic part, and the monopole charge
\begin{align}
\label{eq:rhotot}
  \rho^\mathrm{tot}\left(\mathbf{r}\right)&=\rho^e\left(\mathbf{r}\right)+\rho^i\left(\mathbf{r}\right)+\rho^\mathrm{mono}\left(\mathbf{r}\right),\notag\\
                             &=-n\left(\mathbf{r}\right)+\sum_j Z_j\delta\left(\mathbf{r}-\mathbf{R}_j\right)\notag\\
                             &\quad-n_\mathrm{dop}\,\delta(z-z_\mathrm{mono}),
\end{align}
where $\rho^e\left(\mathbf{r}\right)$, $\rho^i\left(\mathbf{r}\right)$, and $\rho^\mathrm{mono}\left(\mathbf{r}\right)$
are the electron, ion, and monopole charge densities, respectively.
Furthermore, $Z_j$ is the (pseudo) atomic charge of atom $j$ at position $\mathbf{R}_j$, and
the monopole with a total charge of $-n_\mathrm{dop}\,A$ per unit cell is located at $z=z_\mathrm{mono}$.
Within PBC, the interaction energy associated to the presence of the monopole is given by
\begin{align}
\label{eq:emono}
 E_\mathrm{mono} &= -\int_\Omega\rho^\mathrm{tot}\left(\mathbf{r}\right)\,V_\mathrm{mono}\left(\mathbf{r}\right)d^3r,
\end{align}
where $V_\mathrm{mono}$ can be written as
\begin{align}
\label{eq:potential}
 V_\mathrm{mono}\left(\mathbf{r}\right) &= -2\pi\,n_\mathrm{dop}\,\left(-\left|\overline{\mathrm{z}}\right|+\frac{\overline{\mathrm{z}}^2}{L}+\frac{L}{6}\right).
\end{align}
Here $\overline{\mathrm{z}}=z-z_\mathrm{mono}$ with $\overline{\mathrm{z}}\in\left[-L/2;L/2\right]$ measures the distance to the monopole which is at $z_\mathrm{mono}$ and
$L$ is the length of the periodically repeated unit cell along $\mathbf{\hat{z}}$ (see Fig.~\ref{fig:pots}).
Notice that $V_\mathrm{mono}\left(\mathbf{r}\right)$ represents the potential generated in PBC
by the monopole plane of total charge $-n_\mathrm{dop}$ placed at $z_\mathrm{mono}$ (the linear term)
and by a uniform jellium density of opposite total charge $+n_\mathrm{dop}$ (the quadratic term).
This jellium charge cancels
the uniform background charge which is used in standard plane-wave \textit{ab initio}
codes in order to have a neutral system\cite{gava2009}. The last term in Eq.~(\ref{eq:potential}) is constant and it is conventionally chosen in order to
have $\int V_\mathrm{mono}\left(\mathbf{r}\right) d^3r=0$.

To eliminate the electrostatic interactions between repeated cells in the $z$ direction, we introduce a dipole formed by two opposite-charged planes at $z_\mathrm{dip}-d_\mathrm{dip}/2$ and $z_\mathrm{dip}+d_\mathrm{dip}/2$.
The dipole should be placed between the monopole and the vacuum. This condition is realized by choosing
$z_\mathrm{dip}+d_\mathrm{dip}/2=z_\mathrm{mono}-\epsilon$, where $\epsilon$ is a small positive number.
In order to calculate the dipole energy $E_\mathrm{dip}$ we need to determine the
electric dipole moment per unit surface $m$ of the full system (\ie, using the sum of ionic, electronic, and monopole charge density)\cite{bengtsson1999,meyer2001}
\begin{align}
\label{eq:dipole}
 m &= \int_\Omega\left[\frac{f\left(\tilde{z}\right)}{A}\,\rho^\mathrm{tot}\left(\mathbf{r}\right)\right]\,d^3r.
\end{align}
Here $f\left(\tilde{z}\right)$ is a periodic function having different definitions for values of $z$
that are inbetween the two charged planes of the dipole or elsewhere, namely:
\begin{align}
\label{eq:f}
 f\left(\tilde{z}\right) &= \left\{
 \begin{matrix}
  \tilde{z}-L/2                                          &\forall \tilde{z}\geq d_\mathrm{dip}/2\\
  -\tilde{z}\left(L-d_\mathrm{dip}\right)/d_\mathrm{dip} &\forall \tilde{z}<d_\mathrm{dip}/2
 \end{matrix}
 \right.
\end{align}
with 
$\tilde{z}=z-z_\mathrm{dip}$, $\tilde{z}\in\left[-d_\mathrm{dip}/2;L-d_\mathrm{dip}/2\right]$ (and periodically repeated).
The dipole energy is given by\cite{notedipc}
\begin{align}
\label{eq:Edip}
 E_\mathrm{dip} &= -\frac{1}{2}\,\int_\Omega\rho^\mathrm{tot}\left(\mathbf{r}\right)\,V_\mathrm{dip}\left(\mathbf{r}\right)d^3r,
\end{align}
while the potential generated by the two planes of opposite charge can be written as
\begin{align}
\label{eq:dip_pot}
 V_\mathrm{dip}\left(\mathbf{r}\right) &= -\frac{4\pi m}{L}\,f\left(\tilde{z}\right).
\end{align}
Note that in a self-consistent calculation the charge density changes with each iteration. Thus, the dipole moment per unit surface $m$ and the
corresponding potential which represents the dipole have to be recalculated on each iteration until self-consistency is achieved.

Finally, we place a barrier, mimicking the repulsion by the dielectric, to separate the system from the monopole plane. Such barrier, of height $V_0$ and thickness $d_b$,  
starts at $z_1=z_\mathrm{dip}-d_\mathrm{dip}/2$, in correspondence of the first plane of the dipole,
and ends at $z_2=z_\mathrm{dip}-d_\mathrm{dip}/2+d_b$. The associated energy is
\begin{align}
\label{eq:ebarrier}
 E_b &= -\int_\Omega V_b\left(\mathbf{r}\right)\rho^\mathrm{tot}\left(\mathbf{r}\right)d^3r,
\end{align}
where $V_b\left(\mathbf{r}\right)$ is a periodic function of $z$ defined on the interval $z\in[0,L]$ as 
\begin{align}
\label{eq:vbarrier}
 V_b\left(\mathbf{r}\right) &= \left\{
 \begin{matrix}
  V_0,&\forall z\in[z_1,z_2]\\
  0,  &\forall z\notin[z_1,z_2]
 \end{matrix}
\right..
\end{align}
In the present implementation, we have chosen a linear transition from $V_b=0$
to $V_b=V_0$ within $d_\mathrm{dip}$ in order to minimize the fluctuation of the potential in $\mathbf{k}$ space.

The final, full Kohn-Sham potential of the gated system can be written as
\begin{align}
\label{eq:vks}
 V_\mathrm{KS}\left(\mathbf{r}\right)&=\frac{\delta\left[E_\mathrm{tot}-\sum_i\langle\psi_i\left|T\right|\psi_i\rangle\right]}{\delta n\left(\mathbf{r}\right)}\notag\\
   &= V^i_\mathrm{per}\left(\mathbf{r}\right) + V_H\left(\mathbf{r}\right) + V_\mathrm{XC}\left(\mathbf{r}\right)\notag\\
   &\quad + V_\mathrm{mono}\left(\mathbf{r}\right) + V_\mathrm{dip}\left(\mathbf{r}\right) + V_b\left(\mathbf{r}\right),
\end{align}
where $V^i_\mathrm{per}\left(\mathbf{r}\right)$ is the bare, periodic electron-ion potential, $V_H\left(\mathbf{r}\right)$ is the normal Hartree potential
\begin{align}
\label{eq:vh}
 V_H\left(\mathbf{r}\right) &= \int_\Omega\frac{n\left(\mathbf{r}'\right)}{\left|\mathbf{r}-\mathbf{r}'\right|} d^3r',
\end{align}
$V_\mathrm{XC}\left(\mathbf{r}\right)$ is the exchange-correlation potential, and $V_\mathrm{mono}\left(\mathbf{r}\right)$, $V_\mathrm{dip}\left(\mathbf{r}\right)$, $V_b\left(\mathbf{r}\right)$,
are the monopole, dipole, and potential barrier, given in Eqs.~(\ref{eq:potential}), (\ref{eq:dip_pot}), and (\ref{eq:vbarrier}), respectively.
The different parts of the Kohn-Sham potential are also shown in Fig.~\ref{fig:pots}.

For the structural optimization we need to calculate the total force $\mathbf{F}^j$ on atom $j$.
Using the Hellman-Feynman theorem and Eq.~(\ref{eq:etot}) the force can be written as\footnote{Notice that if the barrier
potential $V_b\left(\mathbf{r}\right)$ is correctly chosen, the nuclei never enter the barrier region and the last term
of Eq.(\ref{eq:ftotal}) is always equal to zero. The repulsive force on the system is mediated by the tail of the electron
distribution within the barrier region. Such repulsion is included in the terms of Eq.(\ref{eq:ftotal}) which implicitly
or explicitly depend on the electron charge density $\rho^e\left(\mathbf{r}\right)$.}
\begin{align}
\label{eq:ftotal}
  \mathbf{F}^j &= -\int\rho^e\left(\mathbf{r}\right)\nabla_{\mathbf{R}_j}V^i_\mathrm{per}\left(\mathbf{r}\right)d^3r-\nabla_{\mathbf{R}_j}E_{i-i}\notag\\
               &\quad +Z_j\left(\frac{\partial V_\mathrm{mono}\left(\mathbf{R}_j\right)}{\partial\mathbf{r}}+\frac{\partial V_\mathrm{dip}\left(\mathbf{R}_j\right)}{\partial\mathbf{r}}\right.\notag\\
               &\qquad\qquad\left.+\frac{\partial V_b\left(\mathbf{R}_j\right)}{\partial\mathbf{r}}\right).
\end{align}
Using this equation we can now relax the system in an FET setup. We explicitly verified the force calculation by finite differences.

\subsection{Computational details}
\label{DFT}
All calculations were performed using DFT with a modified version of the {\sc PWscf} code of the {\sc Quantum ESPRESSO} package\cite{quantumespresso}.
We used the Perdew-Zunger local density approximation (LDA) exchange-correlation functional\cite{pz81} together with ultrasoft pseudopotentials including core corrections.
We cross-checked our results with the Perdew-Burke-Ernzerhof (PBE) functional\cite{pbe96} including dispersion corrections\cite{grimme2006} (``+D2'').
A plane-wave basis set was used to describe the valence electron wave function and density up to a kinetic energy cutoff of $45$ and
$500\:\mathrm{Ry}$ ($1\:\mathrm{Ry}\approx13.6\:\mathrm{eV}$), respectively. The electronic eigenstates have been
occupied with a Fermi-Dirac distribution, using an electronic temperature of $300\:\mathrm{K}$.
The BZ integration has been performed with
a Monkhorst-Pack grid\cite{monkhorst1976} of $64\times64\times1$ for the charged systems and $16\times16\times1$ $\mathbf{k}$ points
for the neutral one. In order to correctly determine the Fermi energy in the charged system, we performed a non-self-consistent
calculation on a denser $\mathbf{k}$-point grid of $90\times90\times1$ points starting from the converged charge density.
The self-consistent solution of the Kohn-Sham equations was obtained when the total energy changed by less
than $10^{-9}\:\mathrm{Ry}$ and the maximum force on all atoms was less than $5\cdot10^{-4}\:\mathrm{Ry}\:a_0^{-1}$
($a_0\approx0.529177\:\angstrom$ is the Bohr radius). A tight force convergence threshold was necessary since the force on the atoms due to the
charged plane representing the gate can be as small as $10^{-2}\:\mathrm{Ry}\:a_0^{-1}$ for the lighter nitrogen atoms.

Using the experimental lattice parameters of bulk $\beta$-ZrNCl as starting geometry\cite{shamoto1998}, we relaxed the unit cell for the bulk system until the total stress on the system
was less than $0.1\:\mathrm{kbar}$. The lattice parameters thus determined agree well with the experimental values: for the in-plane lattice
constant we found a value of $a=3.5622\:\angstrom$ (Exp.: $a=3.5974\:\angstrom$) and for the perpendicular constant we calculated
$c=27.035\:\angstrom$ (Exp.: $a=27.548\:\angstrom$). The deviation of about 1\% is in the range of variations typically
found for LDA calculations. Using PBE+D2 the deviations can be decreased to less than 1\%, but the convergence for the layered-2D systems is slower,
especially if ZrNCl is close to the position $z_0$ of the monopole potential.
Since we are mainly interested in the general behavior of the system under the influence of a gate voltage we continue to use the LDA.
We cross-checked our results for selected gate voltages using PBE+D2 and got the same general behavior with only small changes.
The final geometry of the bulk system was used as the starting geometry for the calculations of the layered-2D systems fixing the size of
the unit cell in plane and increasing the perpendicular size such that the vacuum region between the repeated images was at least $20\:\angstrom$.
We investigated single-, double-, and triple-layer systems with different doping levels ranging from $0.05$ to $0.35$ electrons per unit cell
(\ie, $0.025$ to $0.175$ electrons per Zr atom since we are using the hexagonal unit cell) which corresponds to a doping charge per unit surface ranging from
$n_\mathrm{dop}\approx4.55\cdot10^{13}\:\mathrm{cm^{-2}}$ to $n_\mathrm{dop}\approx3.18\cdot10^{14}\:\mathrm{cm^{-2}}$.

The dipole was placed at $z_\mathrm{dip}=d_\mathrm{dip}/2$ with $d_\mathrm{dip}=0.01\:L$ and $L\approx35\:\angstrom$ for single- and double-layer
ZrNCl, and $L\approx46\:\angstrom$ in the trilayer case. A potential barrier with an height of $V_0=1.5\:\mathrm{Ry}$ and a width of $d_b=0.1\:L$ was used
in order to prevent the ions from moving too close to the monopole. The final results were found to be independent of the separation of the dipole planes,
as well as the barrier height and width as long as it is high or thick enough to ensure that $\rho^e\left(z_\mathrm{mono}\right)=\rho^e\left(z_\mathrm{dip}\right)=0$.

\section{Results}
\subsection{Structural changes under electrochemical doping}
\label{geometry}
In the relevant experiment of Ref.~\citenum{ye2010} on ZrNCl an ionic liquid was used as dielectric and the gate voltage $V_G$ was applied between the ionic liquid and ZrNCl.
The mobile cation and anion of the ionic liquid move towards the oppositely charged electrode which in turn leads to the
formation of an electric double layer (EDL) at the interface to ZrNCl. The EDL at the ZrNCl/ionic liquid interface acts as
a capacitor with a huge capacitance (``supercapacitor'') and can accumulate charges in the transport channel between source and drain. Using
such an electric double-layer transistor one can achieve large carrier densities up to $n_\mathrm{dop}=8\times10^{14}\,\mathrm{cm^{-2}}$ (see Ref.~\citenum{yuan2009})
thus allowing the investigation of the transition from a band insulator to a metal and, eventually, to a superconductor\cite{ye2010}.
\begin{figure}[b]
 \centering
  \includegraphics[width=0.125\textwidth,clip=]{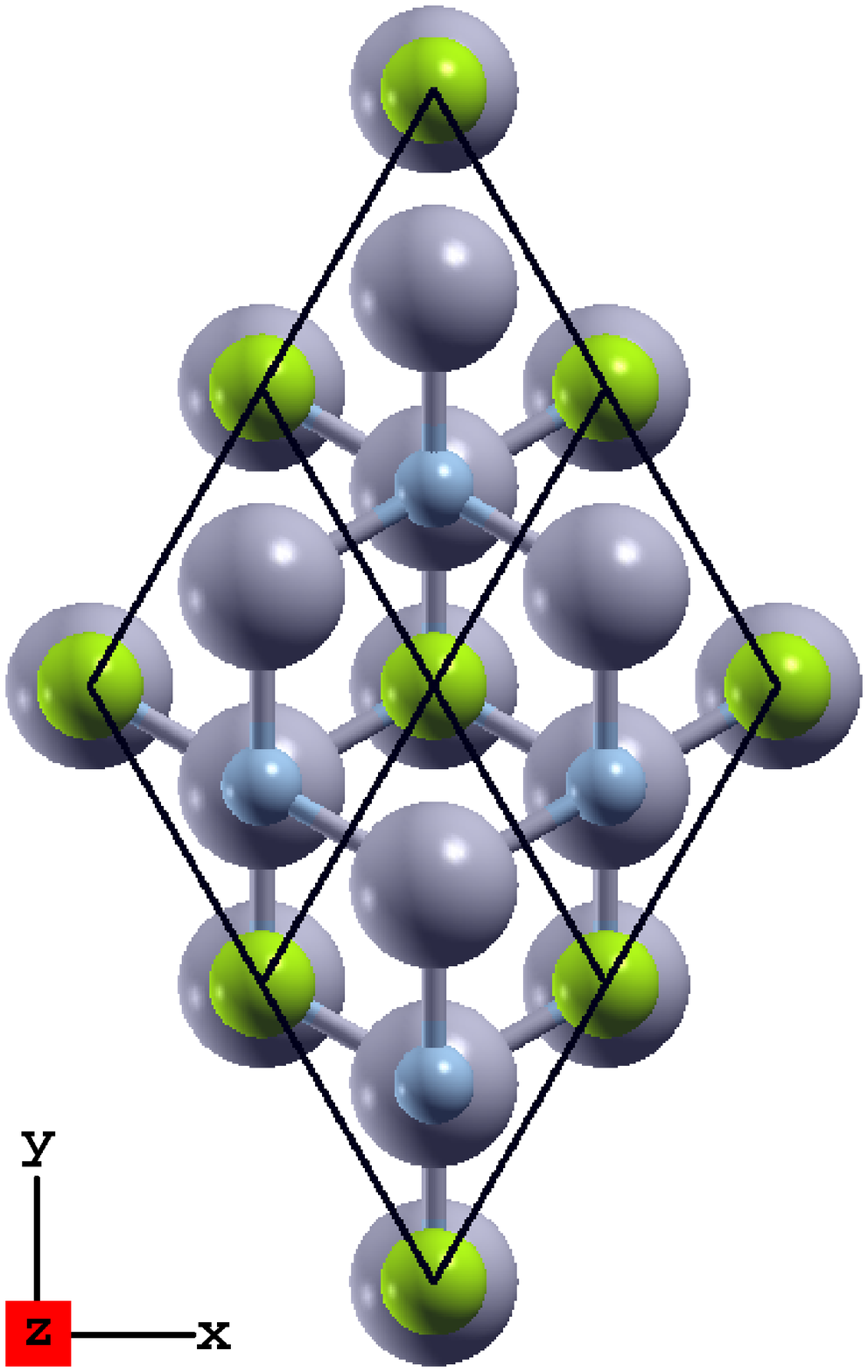}
  \includegraphics[width=0.3\textwidth,clip=]{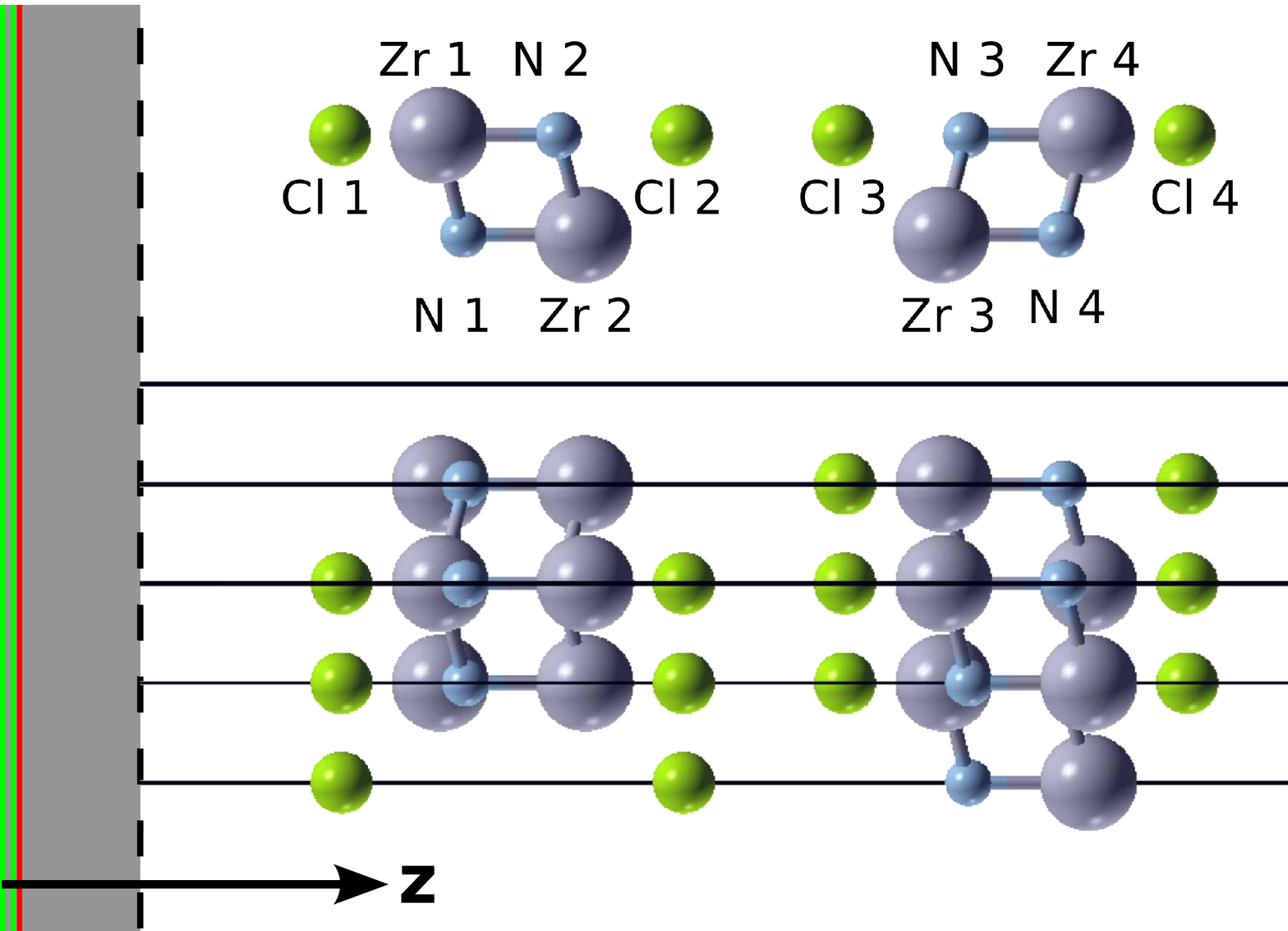}
 \caption{\label{fig:distdef}(color online) Structure of the ZrNCl double-layer system with the unit cell indicated by solid, black lines.
          Green balls -- Chlorine; large, gray balls -- Zirconium; small, bluish-gray balls -- Nitrogen.
          In the on-top view on the left-hand side one can clearly see the (buckled) honeycomb lattice formed by the nitrogen and zirconium atoms.}
\end{figure}
The thickness of the EDL is mainly determined by the size of the ions, the microscopic
structure of the surface, and the interaction between both. In this work we simulate the gate with the monopole potential (Eq.~(\ref{eq:potential}))
as shown in Fig.~\ref{fig:pots}, which has a fixed position $z_\mathrm{mono}$. We thus need to relax the nanolayers towards the charged plane of the monopole
potential until the attractive force between the two oppositely charged systems is compensated by the repulsive force due to the electron
density within the potential barrier, $V_b$. In this section we
analyze the final geometry with increasing doping (\ie, increasing strength of the monopole potential) in detail.
Figure \ref{fig:distdef} shows the atomic structure of a double-layer ZrNCl together with the definition of the atoms which are used in
the following discussion and their position with respect to the interface between system and ionic liquid.
\begin{figure}
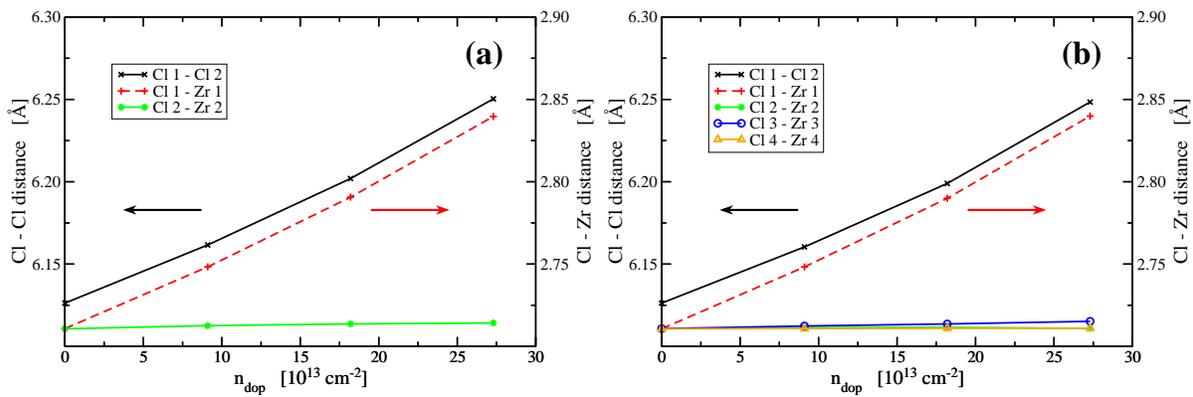

 \centering
 \includegraphics[scale=0.29,clip=]{figure4a.eps}
 \includegraphics[scale=0.29,clip=]{figure4b.eps}
 \caption{\label{fig:Cldist} (color online) Geometrical changes with increasing gate-induced doping for a single layer (a) and a double layer (b)
          of ZrNCl (\cf, Fig.~\ref{fig:distdef} for the definition of the different atoms). The largest change can be
          seen in the Cl-Zr distance for the chlorine closest to the ionic liquid (``Cl 1 - Zr 1''), while the other distances are nearly unchanged.
          Furthermore, the differences in the Cl-Zr distances between a single layer of ZrNCl and the bilayer system are negligible.
          The bulk value for the Cl 1 - Cl 2 distance is $6.127\:\angstrom$ while the Cl-Zr bond length is $2.711\:\angstrom$.}
\end{figure}

The bonding between the neutral layers of ZrNCl is very weak\cite{heid2005,akashi2012} which allows for the intercalation with alkaline
elements or even molecules\cite{kasahara2010}. Thus, the interaction between ZrNCl and the monopole potential representing the gate could
in principle strongly change the interlayer distance depending on the distribution of the doping charge.
In our calculations however, we find that the interlayer distance is nearly unchanged. Instead the thickness of the nanolayers system
is increased, due mainly to the change in the Cl-Zr distance for the chlorine closest to the gate, while the other distances are
nearly constant as can be seen in Fig.~\ref{fig:Cldist}. 

The interlayer distance in the bilayer case changes by less than 1\% compared to the value of bulk ZrNCl.
Triple-layer ZrNCl shows the same general behavior---we find an increase of about 1\% in the distance between the middle layer
and the one next to the charged plane from $3.537\:\angstrom$ for $n_\mathrm{dop}\approx9\cdot10^{13}\:\mathrm{cm^{-2}}$ to $3.576\:\angstrom$ for
$n_\mathrm{dop}\approx27\cdot10^{13}\:\mathrm{cm^{-2}}$ (bulk value $3.543\:\angstrom$).
The absolute value of these interlayer distances, in principle, depends on the correct description of the van der Waals interactions
between the layers and, thus, might be questionable. The most important result is that for all investigated
systems the major changes in the geometry occur in the layer next to the monopole (see Fig.~\ref{fig:distdef}).

\begin{figure}
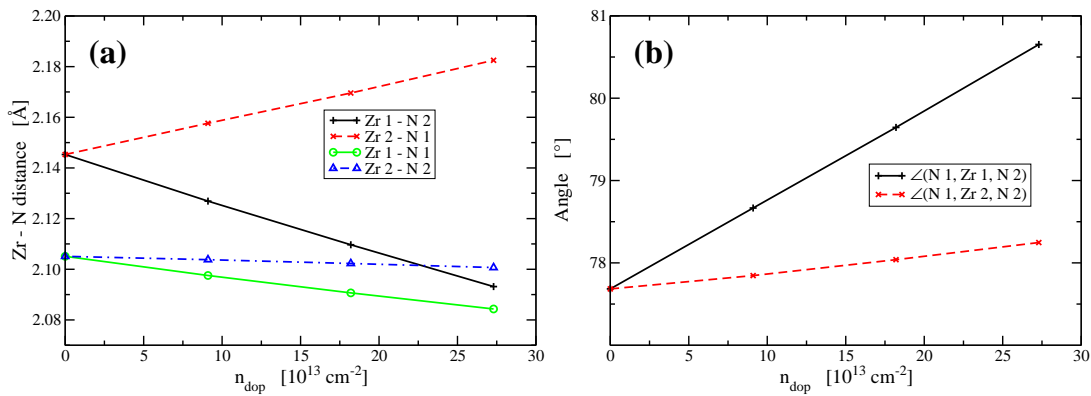

 \centering
 \includegraphics[scale=0.29,clip=]{figure5a.eps}
 \includegraphics[scale=0.29,clip=]{figure5b.eps}
 \caption{\label{fig:ZrNdist_angles}(color online) (a) Zr-N bond lengths for a single-layer of ZrNCl and (b) the angles between the
          different nitrogen and zirconium atoms as defined in Fig.~\ref{fig:distdef}.
          Again there are no differences between different number of layers which is why we here only show the results for the single-layer case.
          The bulk values for the in-plane and the out-of-plane bonds are $2.105\:\angstrom$ and
          $2.146\:\angstrom$, respectively. The bulk value for the angles is $77.69^\circ$.}
\end{figure}

This is also the case for the different Zr-N bond lengths, and angles of the Zr-N rhomboid. In the double- and triple-layer system
relevant changes occur only in the layer next to the monopole.
As the relaxed geometry of the ZrNCl layer closer to the monopole are very similar for the single-, double-, and triple-layer case,
we only present the results of the structural minimization of a single layer (see Fig.~\ref{fig:ZrNdist_angles}).
Due to the large difference between the electronegativity of
nitrogen and zirconium, the nitrogen is partially negative. Accordingly, N is attracted
by the positively charged plane representing the gate. This leads to an increase of the bond length labeled ``Zr 2 - N 1'' which is
the bond with the nitrogen closer to the monopole as can be seen in Fig.~\ref{fig:ZrNdist_angles}(a). On the other hand, the bond length in which
the zirconium is closer to the monopole is decreased (``Zr 1 - N 2''). Since the in-plane bonds are nearly unchanged
the N-Zr-N angles are modified considerably as can be seen in Fig.~\ref{fig:ZrNdist_angles}(b).
These changes in the geometry of the layer next to the monopole are due to the fact that the doping charge is mainly localized within this
layer as we will see in the next section.

\subsection{Electronic structure}
As shown in the last section, the structure of the layer closest to the monopole is strongly affected by the electric field.
This is due to the interplay between the electric field, the doping charge, and the specific band structure of ZrNCl.
The conduction band
minimum is located at the $\mathrm{K}$ and $\mathrm{K}'$ point\cite{akashi2012,sugimoto2004,heid2005} and has a nearly quadratic in-plane dispersion $E(\mathbf{k})\propto k^2_x+k^2_y$. Furthermore, the band
consists mainly of Zr $d$-states which together with the weak interlayer interaction leads to a negligible
dispersion along $k_z$. Thus, it can be expected that the doping charge is mainly localized within the center
of the layers closest to the ionic liquid.
\begin{figure}[b]
 \centering
 \includegraphics[scale=0.258,clip=]{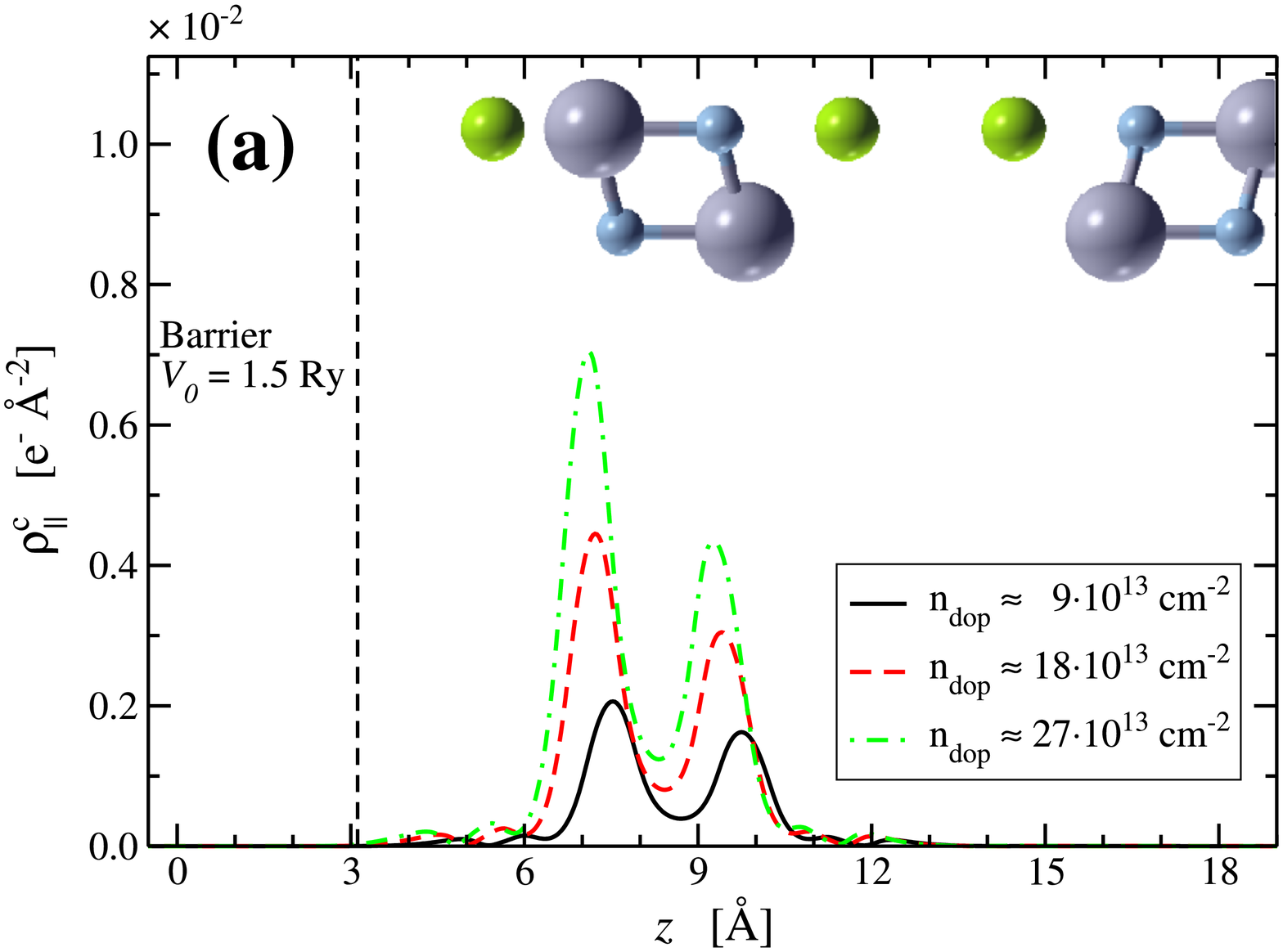}
 \includegraphics[scale=0.258,clip=]{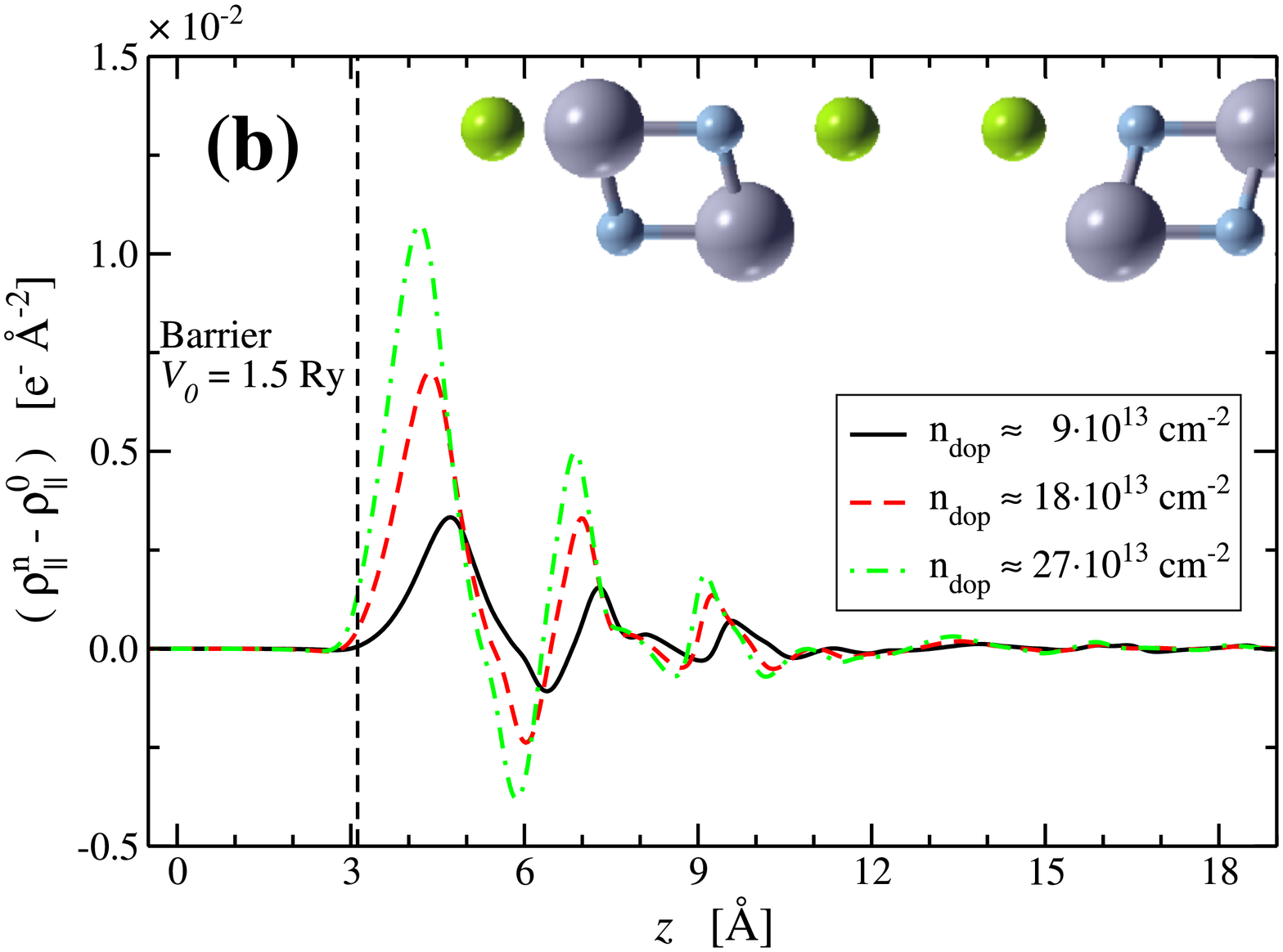}
 \caption{\label{fig:charge_ildos}(color online) (a) Distribution of the conduction band electrons and (b) difference between the full charge
          density of the doped system $\rho^n_{||}$ and the undoped system $\rho^0_{||}$ for bilayer ZrNCl as a function of $z$.
          The doping occurs mainly in the layer which is closer to the ionic liquid as can be seen in (a) showing the planar average of the
          charge density of the electrons in the conduction band $\rho^c_{||}$.
          In panel (b) the planar averaged difference between the full charge density of the doped system $\rho^n_{||}$ and the undoped
          system $\rho^0_{||}$ is shown. The density of the uncharged system $\rho^0_{||}$ was calculated with the geometry of the charged
          one.}
\end{figure}
Figure \ref{fig:charge_ildos}(a) shows the distribution of the conduction band electrons
in bilayer ZrNCl averaged over the directions parallel to the layers for three doping levels ranging from
$n_\mathrm{dop}\approx9\cdot10^{13}\:\mathrm{cm^{-2}}$ to $n_\mathrm{dop}\approx27\cdot10^{13}\:\mathrm{cm^{-2}}$.
The doping charge occupying the conduction band is mainly localized in the center of the layer closest to the monopole
and has a strong asymmetry between the two Zr-N planes on either side.
This asymmetry is due to the electric field and the induced structural changes. The field induce not only
an asymmetry in the geometry and the distribution of the conduction band electrons but also an increased
surface dipole $m$. In Fig.~\ref{fig:charge_ildos}(b) the difference between the full
charge density of the doped system $\rho^n_{||}$ and the undoped system $\rho^0_{||}$ is shown, where $\rho^0_{||}$
was calculated with the relaxed geometry of the charged one. This charge difference $\left(\rho^n_{||}-\rho^0_{||}\right)$
represents the screening charge of the gate electric field, including both the metallic screening of the conduction
band and the dielectric screening of the lower, occupied valence bands which rearrange under the applied electric
field.
The field has two main effects as can be seen in Fig.~\ref{fig:charge_ildos}:
firstly, it leads to an asymmetry in the charge distribution and consequently in the geometry as discussed in the
last section, and secondly, it increases the surface dipole---not only due to the structural changes but also
due to the rearrangement of the charge density of the undoped system $\rho^0_{||}$.
In fact, the induced dipole is largest close to the last chlorine (Fig.~\ref{fig:charge_ildos}(b)), even if the doping charge in the conduction band
is localized within the center of the layer (Fig.~\ref{fig:charge_ildos}(a)).

\begin{figure}
 \centering
 \includegraphics[scale=0.258,clip=]{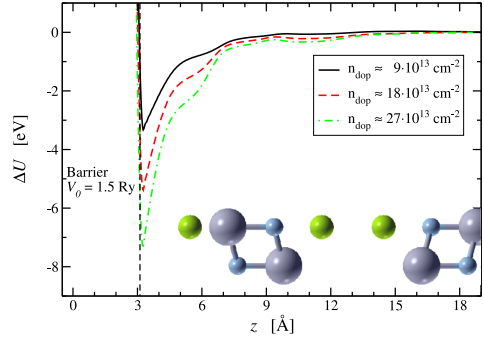}
 \caption{\label{fig:pot}(color online) Difference between the the planar averaged potential $U\left(\mathbf{r}\right)=V_\mathrm{KS}\left(\mathbf{r}\right)-V_\mathrm{XC}\left(\mathbf{r}\right)$
          for the charged and the uncharged system as function of $z$.
          The derivative $\partial\Delta U\left(\mathbf{r}\right)/\partial z$ is proportional to the effective electric field $E_z$.}
\end{figure}
The strong localization of the doping charge within the first layer also leads to a complete screening of the electric field.
Figure \ref{fig:pot} shows the difference of the planar averaged potential $U\left(\mathbf{r}\right)=V_\mathrm{KS}\left(\mathbf{r}\right)-V_\mathrm{XC}\left(\mathbf{r}\right)$ of the charged and the uncharged system $\Delta U\left(\mathbf{r}\right)=U^n\left(\mathbf{r}\right)-U^0\left(\mathbf{r}\right)$, the derivative of which
being proportional to the effective electric field $\partial\Delta U\left(\mathbf{r}\right)/\partial z\propto E_z$.
The slope of the potential close to the potential barrier at $z\approx3\:\angstrom$ increases with increasing doping and the total potential changes mainly
within the layer of ZrNCl which is closer to the monopole. The largest change and, thus, the largest electric field can be found close to the outermost chlorine atom.

\begin{figure}
 \centering
 \includegraphics[scale=0.3,clip=]{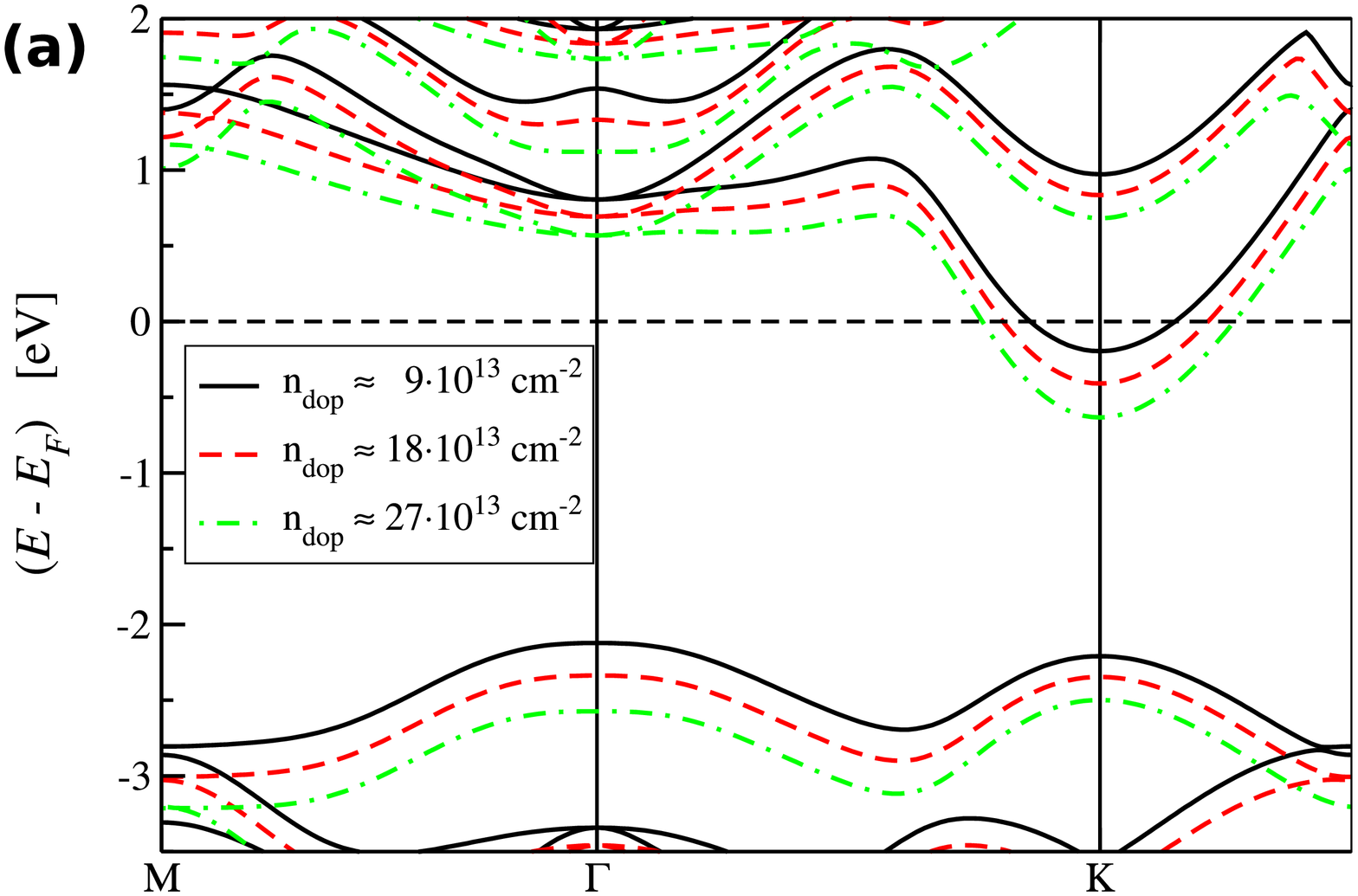}
 \includegraphics[scale=0.3,clip=]{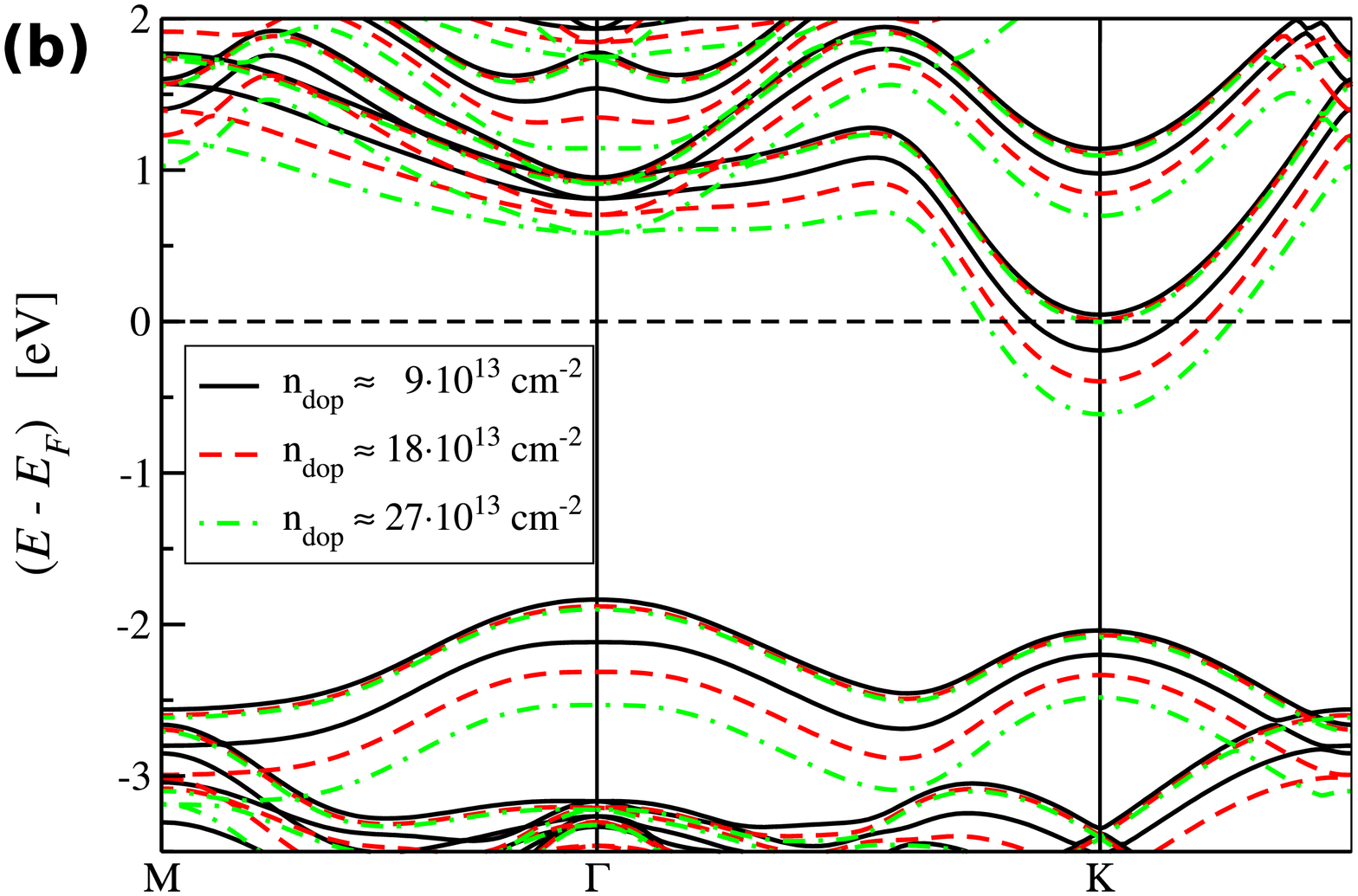}
 \includegraphics[scale=0.3,clip=]{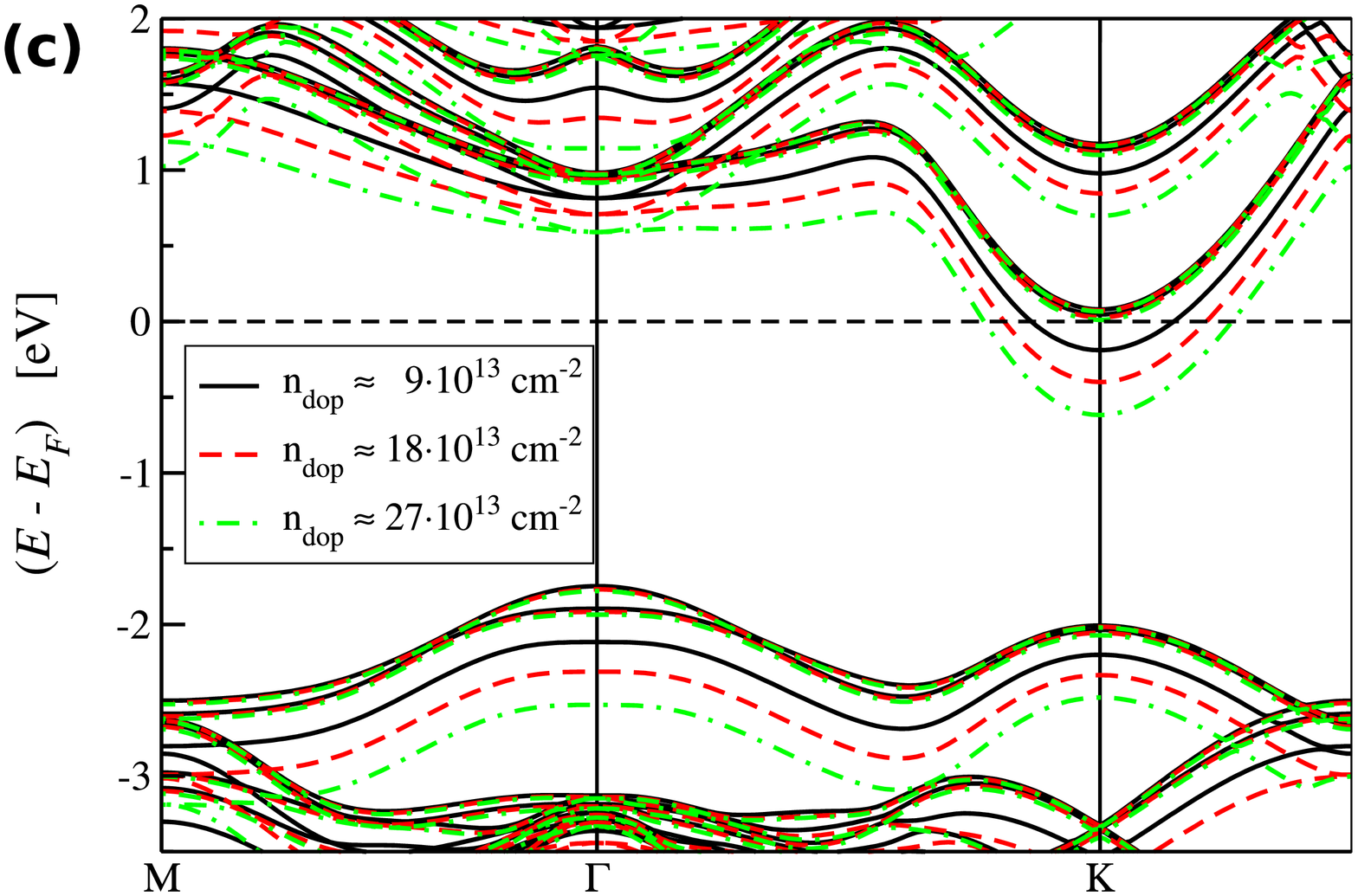}
 \caption{\label{fig:bands}(color online) Bandstructure of (a) monolayer, (b) bilayer, and (c) trilayer ZrNCl with increasing doping.
          The electrons occupy the conduction band minimum at the $\mathrm{K}$ point. For the multi-layer systems this leads to a lifting of the degeneracy between the
          bands localized on the different layers since the doping occurs mainly in the layer which is closer to the monopole (\cf, Fig.~\ref{fig:charge_ildos}).
          The valance band is unchanged despite the geometrical changes (\cf, Figs.~\ref{fig:Cldist} and
          \ref{fig:ZrNdist_angles}).}
\end{figure}

\begin{figure}
 \centering
 \includegraphics[scale=0.3,clip=]{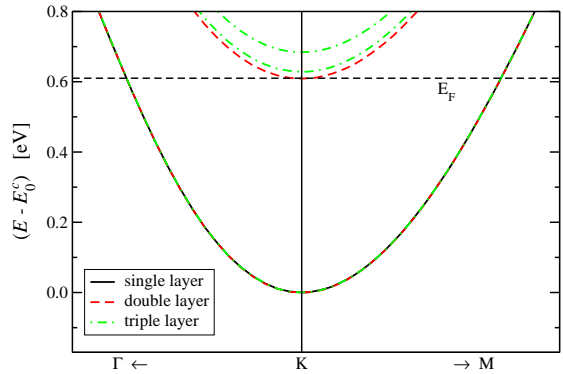}
 \caption{\label{fig:bands_close}(color online). Comparison of the band structure near the $\mathrm{K}$ point for different number of ZrNCl layers
          with the minimum of the conduction band $E^c_0$ set to zero.
          The occupied band has exactly the same dispersion for all systems near the $\mathrm{K}$ point.
          The Fermi energy (dashed black line) agrees within the precision of our calculations and is at $(E-E^c_0)=0.610\pm0.020\:\mathrm{eV}$.}
\end{figure}

The band structure for different number of layers is shown in Fig.~\ref{fig:bands}.
Since the doping charge is localized within the layer close to the ionic liquid/system interface, only the electronic structure of this layer is influenced and, accordingly,
the degeneracy of the conduction band at the $\mathrm{K}$ and $\mathrm{K}'$ point is lifted.
The bands localized on the doped layer are shifted down in energy while those
of the undoped layers are unchanged as can be seen in Figs.~\ref{fig:bands}(b) and (c) for bilayer and trilayer ZrNCl, respectively.
Also the former valence band is nearly unchanged despite the geometrical changes (\cf, Figs.~\ref{fig:Cldist} and \ref{fig:ZrNdist_angles}).
Aligning the band structure for different number of layers to the minimum of the conduction band $E^c_0$ as in Fig.~\ref{fig:bands_close} demonstrates again that there is effectively no
difference between the monolayer/bilayer/trilayer systems. Thus, in this section we focus in the following on the single-layer system and analyze
the variations of the partially occupied band on doping in more detail.

\begin{figure}
 \centering
 \includegraphics[scale=0.3,clip=]{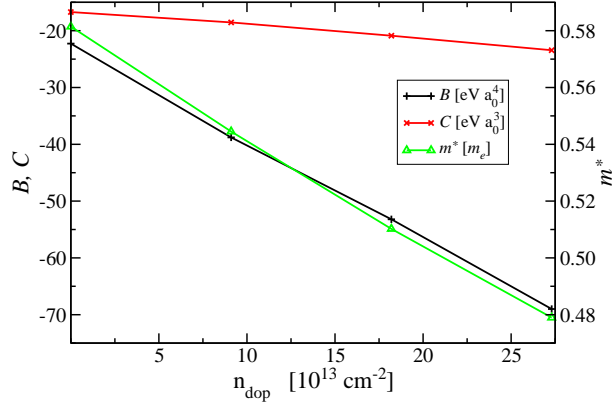}
 \caption{\label{fig:parameters}(color online). Parameters $m^*$, $B$, and $C$ obtained by fitting Eq.~(\ref{eq:dispersion}) to the band structure of single-layer ZrNCl
          for three different doping levels. The effective mass $m^*$ is given in atomic units, \ie, in units of the electron mass $m_e$. $B$ and $C$ are in units of
          $\mathrm{eV\,a_0^4}$ and $\mathrm{eV\,a_0^3}$, respectively.}
\end{figure}

Charging of the monolayer does not simply act as a rigid doping as it also affects the shape of the band.
In order to describe the changes in the band structure
quantitatively we model the dispersion relation $E(\mathbf{k})$ near the $\mathrm{K}$ and $\mathrm{K}'$ points by a
4\textsuperscript{th} order Taylor expansion in $\mathbf{k}$, compatible with the symmetry of the bands:
\begin{align}
\label{eq:dispersion}
 E\left(\mathbf{k}\right) &= E^c_0 + \frac{\hbar^2}{2\,m^*}\,\left(k_x^2 + k_y^2\right) + B\,\left(k_x^2 + k_y^2\right)^2\notag\\
                          &\quad+ C\,\left(k_x^3 - 3\,k_x\,k_y^2\right).
\end{align}
Here $k_x$ is aligned with the $\Gamma\rightarrow\mathrm{K}$ direction and $|\mathbf{k}|=k$ measures the distance
from the $\mathrm{K}$, $\mathrm{K}'$ special points.
The $k^3$ term is needed to describe the trigonal warping of the conduction band while both the 3\textsuperscript{rd} the 4\textsuperscript{th} order term
are necessary to allow for the deviations of the density of states from the constant energy dependence of a perfect 2D electron gas.
The parameters $m^*$, $B$, and $C$ obtained by fitting Eq.~(\ref{eq:dispersion}) to the band structure of single-layer ZrNCl for three different doping levels
are depicted in Fig.~\ref{fig:parameters}. 
Most interestingly, the effective mass $m^*$ changes by 20\% in the doping
regime investigated here, which would lead to
the same change in the density of states (DOS) for a perfect 2D electron gas which can be modeled using just the first two terms
in Eq.~(\ref{eq:dispersion}).

\begin{figure}
 \centering
 \includegraphics[scale=0.3,clip=]{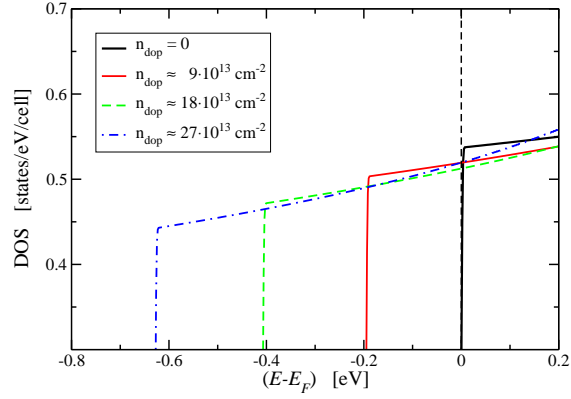}
 \caption{\label{fig:dos}(color online) DOS per unit cell for single-layer ZrNCl. The DOS was calculated using the dispersion relation in Eq.~(\ref{eq:dispersion})
          and the parameters obtained by fitting it to the band structure near the $\mathrm{K}$ point (Fig.~\ref{fig:parameters}). For the zero charge case the
          minimum of the conduction band was used as Fermi energy $E_F$.
          The first-principles DOS agrees with that calculated using the dispersion relation shown here.}
\end{figure}

Yet the parameters $B$ and $C$, which describe the deviations from a quadratic dispersion for higher values of $k$, change considerably when the doping is increased.
The decrease from about $-22\:\mathrm{eV\,a_0^4}$ to $-69\:\mathrm{eV\,a_0^4}$ for $B$ and from about $-17\:\mathrm{eV\,a_0^3}$ to $-24\:\mathrm{eV\,a_0^3}$ for $C$
compensate the change in the effective mass $m^*$. Accordingly the DOS at the Fermi energy stays more or less constant as can be seen in Fig.~\ref{fig:dos}.
In fact, specific heat measurements on Li intercalated ZrNCl suggested that the DOS at the Fermi energy is not increased with increasing doping\cite{kasahara2009},
which cannot be explained assuming a rigid doping since the DOS is not flat (\cf, Refs.~\citenum{sugimoto2004,heid2005,akashi2012} and Fig.~\ref{fig:dos}).

\begin{figure}
 \centering
 \includegraphics[scale=0.3,clip=]{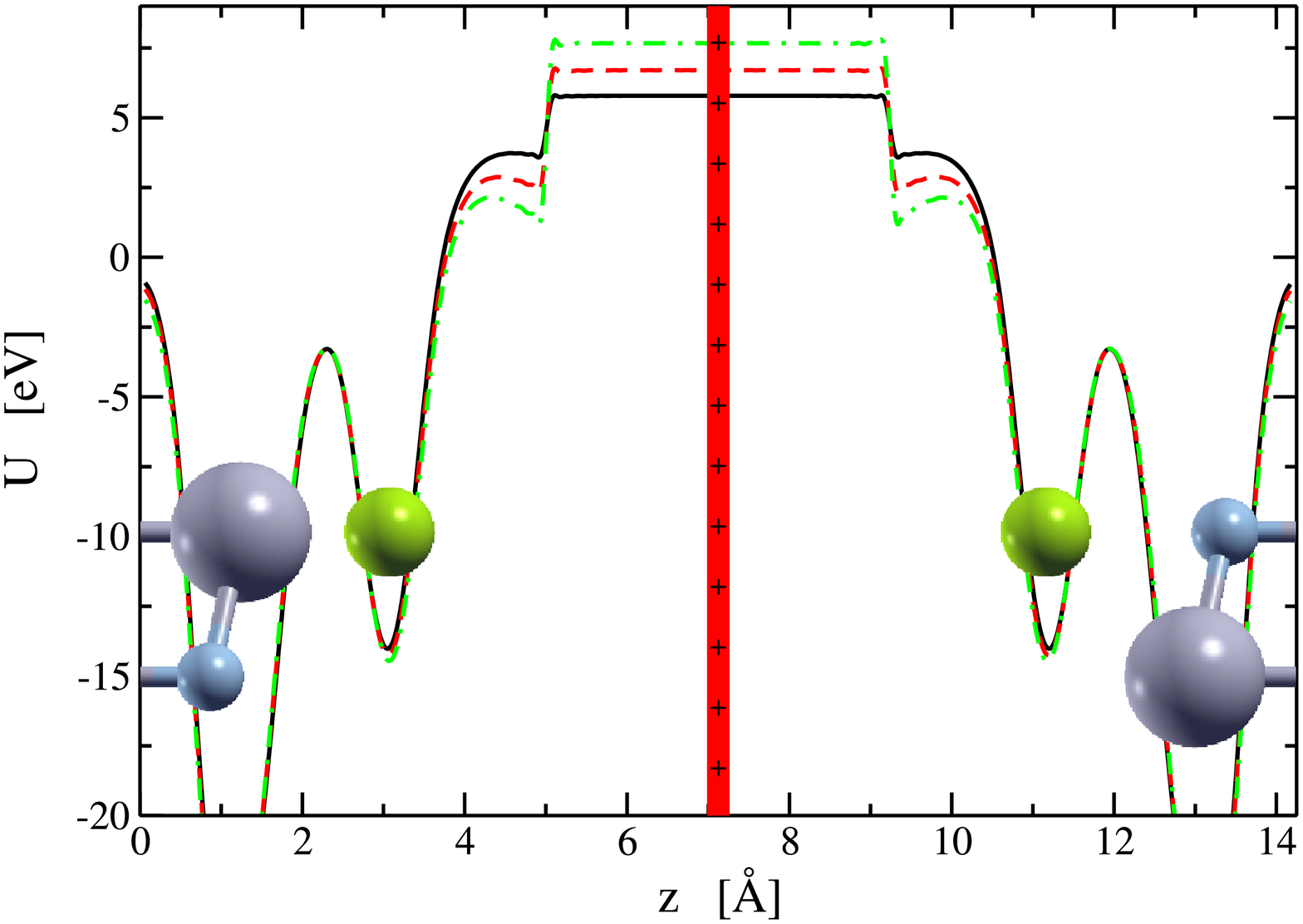}
 \caption{\label{fig:symmetricmodel}(color online) Sketch of the unit cell along the $z$ axis of the symmetric model for the Li intercalation in bulk ZrNCl
          together with the potential $U\left(\mathbf{r}\right)=V_\mathrm{KS}\left(\mathbf{r}\right)-V_\mathrm{XC}\left(\mathbf{r}\right)$ for the 3
          doping concentrations used throughout the paper. The monopole at $z_\mathrm{mono}=L/2$ is indicated with a thick red line.}
\end{figure}

\begin{figure}
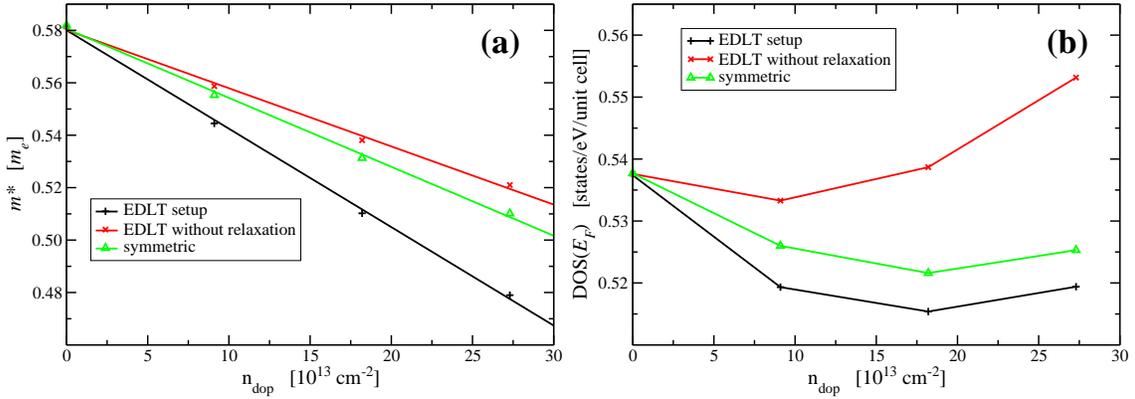

 \centering
 \includegraphics[scale=0.3,clip=]{figure13a.eps}
 \includegraphics[scale=0.3,clip=]{figure13b.eps}
 \caption{\label{fig:mass}(color online) The effective mass $m^*$ obtained by fitting the dispersion relation given in Eq.~(\ref{eq:dispersion}) to the band structure near
          the $\mathrm{K}$ point is shown in (a). The black curve (``EDLT setup'') is the same as in Fig.~\ref{fig:parameters} while the red one
          (gray, ``EDLT without relaxation'') shows the effective mass for ZrNCl which was not relaxed (\ie, in the bulk geometry). The green curve
          (light gray, ``symmetric'') depicts the effective mass for an approximate bulk Li$_x$ZrNCl system in which the Li atoms were represented by Eq.~(\ref{eq:potential})
          (\ie, no homogeneous background charge). Panel (b) shows the corresponding DOS at the Fermi energy.}
\end{figure}

In our simulated electric double-layer transistor (EDLT) the DOS at the Fermi energy is constant since the band structure is changed upon doping.
However, does this picture hold for bulk Li$_x$ZrNCl?
As a first attempt to simulate the Li intercalation in the bulk system with our model, we substitute the Li$_x$ plane
in bulk Li$_x$ZrNCl (\ie, periodic in all three dimension) with a charged plane as described in Eq.~(\ref{eq:potential})---a sketch of the model is shown in Fig.~\ref{fig:symmetricmodel}.
Similarly to the EDLT setup we find a decreasing effective mass even if the decrease is smaller
as can be seen as green (light gray) line in Fig.~\ref{fig:mass}(a) labeled ``symmetric''. Since a rigid structure in field-effect configuration also shows
this behavior (red (grey) curve in Fig.~\ref{fig:mass}(a) labeled ``EDLT without relaxation''), it can neither be related to the EDLT setup
nor to the field-induced structural changes. Instead, it might be a system-specific property.
However, the decreasing effective mass is again compensated by the change in the parameters $B$ and $C$ as can be seen in Fig.~\ref{fig:mass}(b).
Further increasing the doping would ultimately lead to an increasing DOS at the Fermi energy for all investigated systems.
Finally, we also note that the maximum possible doping is smaller if the nanolayer system is not relaxed in the presence of the
electric field generated by the potential in Eq.~(\ref{eq:potential}). This shows again the importance of the structural relaxation
for the correct theoretical description of ZrNCl in an FET setup.

\section{Implications for superconductivity}

\begin{figure}
 \centering
 \includegraphics[width=0.45\textwidth,clip=]{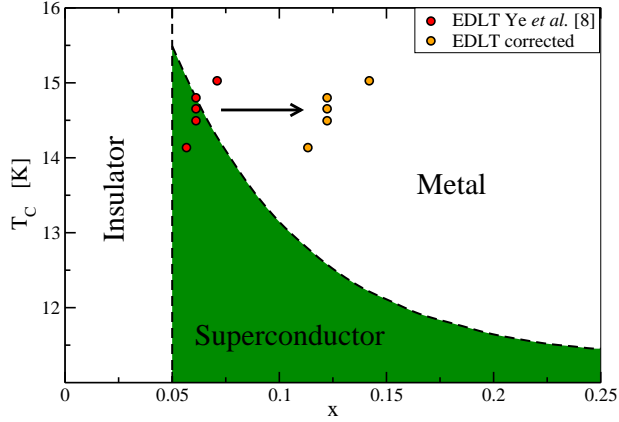}
 \caption{\label{fig:phase_diagram}(color online) Sketch of the unified phase diagram of bulk Li$_\mathrm{x}$ZrNCl and EDLT
          (adapted from Ref.~\citenum{ye2010}).
          Shown is the region close to the metal-insulator transition of bulk Li$_\mathrm{x}$ZrNCl and some of the data points for the liquid gated
          samples\cite{ye2010}. Assuming the correct number of doped layers, the points for the liquid gated samples are shifted towards higher carrier
          concentration (\ie, more electrons per unit cell x). This indicates that the mechanism leading to superconductivity in the EDLT might be different
          from the one in bulk Li$_\mathrm{x}$ZrNCl.}
\end{figure}

Electrochemical doping of ZrNCl is appealing as an insulator-superconductor
transition occurs as a function of charging\cite{ye2010}, as shown in
Fig.~\ref{fig:phase_diagram}. 
The critical doping of the insulator-superconductor transition and 
the behavior of the superconducting critical temperature as a
function of doping depends crucially on the depth of the induced charge layer in the sample.
Indeed, if the induced charge layer is deeper, then for a given
gate voltage the charge density per layer (and thus the
doping) is smaller. Thus knowledge of the charge distribution in the ZrNCl flakes is a crucial
parameter to determine the phase diagram.

While the gate voltage and the total induced charge (Hall effect) are accessible
quantities, the depth of the induced charge layer is not. Thus
experimentalist are forced to use simplified screening models to
access this quantity. In the case of electrochemically doped ZrNCl,
Ye \etal{} in Ref.~\citenum{ye2010} assumed that the induced charge
is localized on two ZrNCl layers. Under this assumption, they obtained the
phase diagram in Fig.~\ref{fig:phase_diagram}, showing a similar
behavior for Li-intercalated and electrochemically doped ZrNCl.
However, we have shown in the last section that this assumption is not well grounded,
as the depth of the induced charge layer is only one ZrNCl
layer, and not two. Furthermore
we have shown that the charge distribution inside the outermost ZrNCl
layer is strongly inhomogeneous.
These findings imply a redrawing of the ZrNCl
phase diagram, as shown in Fig.~\ref{fig:phase_diagram}.
In particular, in the revised phase diagram of electrochemically 
doped samples, superconductivity
occurs at doping larger then $0.12$ and T$_c$ {\it increases}
monotonically from a minimum value of $14\:\mathrm{K}$ up to a maximum of
$\approx 15\:\mathrm{K}$ at a doping of $\approx 0.15$.
This is in striking contrast with chemically
intercalated bulk ZrNCl samples, in which superconductivity occurs at
a doping of $0.05$ with  T$_c=15.5\:\mathrm{K}$. Moreover, in the bulk case, 
T$_c$ {\it decreases} monotonically as a function of doping,
as show in Fig.~\ref{fig:phase_diagram}. Thus, our work 
points out crucial differences in superconductivity in Li intercalated bulk
ZrNCl. Further calculations for, \eg, the transition temperature
are beyond the scope of this paper and will be addressed in a separate work.

\section{Conclusions}

In this work we developed a first principles approach to
field-effect doping. The method allows for the calculation of electronic structure
and structural optimization under an applied electric field in an FET configuration.
We have applied the method to the transition-metal chloronitride ZrNCl and have shown
that the electric field induces substantial deformation
of bond lengths and angles in the outermost ZrNCl layer in contact
with the ionic liquid. The most evident deformation generated by the
electric field is the large change of the the N-Zr-N angles. The
electronic structure is also affected by the
large electric field as the effective mass of the electron is
substantially reduced as a function of increasing charge. The density
of states at the Fermi level is, however, essentially independent on doping. 
Furthermore, we have shown that the depth of the induced charge layer is
only one ZrNCl layer and that the charge distribution inside the outermost ZrNCl
layer is strongly inhomogeneous. These findings imply a substantial revision of
the phase diagram of electrochemically doped ZrNCl and elucidate crucial
differences with superconductivity in Li intercalated bulk ZrNCl.

\begin{acknowledgments}
The authors acknowledge financial support of the Graphene Flagship and of the French National ANR funds
within the \textit{Investissements d'Avenir programme} under reference ANR-11-IDEX-0004-02, ANR-11-BS04-0019
and ANR-13-IS10-0003-01. Computer facilities were provided by CINES, CCRT and IDRIS (project no. x2014091202).
\end{acknowledgments}

\bibliography{ZrNCl}

\end{document}